\documentclass[12pt]{article}

\usepackage[square, sort, numbers, compress]{natbib}

\usepackage{times}

\usepackage{graphicx}
\usepackage{dcolumn}
\usepackage{bm}
\usepackage{amsmath}
\usepackage[utf8]{inputenc}
\usepackage[T1]{fontenc}
\usepackage{lmodern} 
\usepackage[font=small,labelfont=bf,tableposition=top]{caption}
\DeclareCaptionLabelFormat{andtable}{#1~#2  \&  \tablename~\thetable}
\usepackage{subcaption}
\usepackage{multirow}
\usepackage[usenames,dvipsnames]{xcolor}
\usepackage{amsfonts}

\usepackage{xcolor}

\newenvironment{sciabstract}{%
\begin{quote} \bf}
{\end{quote}}

\newcounter{lastnote}

\title{Cyclic transformation of \\ orbital angular momentum modes}

\author{Florian Schlederer,$^{1,\dagger}$ Mario Krenn,$^{1,2,\ast}$ \\ Robert Fickler,$^{1,2,\ddagger}$ Mehul Malik,$^{1,2}$ Anton Zeilinger$^{1,2,\ast}$\\
\\
\normalsize{$^{1}$Institute for Quantum Optics and Quantum Information,}\\
\normalsize{Boltzmanngasse 3, A-1090 Wien, Austria.}\\
\normalsize{$^{\dagger}$present address: National Institute of Informatics, Hitotsubashi 2-1-2,} \\
\normalsize{Chiyoda-ku, Tokyo 101-8403, Japan.}\\
\normalsize{$^{2}$Vienna Center for Quantum Science and Technology (VCQ),}\\
\normalsize{Faculty of Physics, University of Vienna, Boltzmanngasse 5, A-1090 Vienna, Austria.}\\
\normalsize{$^{\ddagger}$present address: Department of Physics and Max Planck Centre for}\\
\normalsize{Extreme and Quantum Photonics, University of Ottawa,Ottawa, K1N 6N5, Canada.}\\
\\
\normalsize{$^\ast$To whom correspondence should be addressed:}\\
\normalsize{mario.krenn@univie.ac.at \&  anton.zeilinger@univie.ac.at.}
}

\date{}

\begin{document} 


\baselineskip24pt


\maketitle


\begin{sciabstract}
The spatial modes of photons are one realization of a QuDit, a quantum system that is described in a D-dimensional Hilbert space. In order to perform quantum information tasks with QuDits, a general class of D-dimensional unitary transformations is needed. Among these, cyclic transformations are an important special case required in many high-dimensional quantum communication protocols. In this paper, we experimentally demonstrate a cyclic transformation in the high-dimensional space of photonic orbital angular momentum (OAM). Using simple linear optical components, we show a successful four-fold cyclic transformation of OAM modes. Interestingly, our experimental setup was found by a computer algorithm. In addition to the four-cyclic transformation, the algorithm also found extensions to higher-dimensional cycles in a hybrid space of OAM and polarization. Besides being useful for quantum cryptography with QuDits, cyclic transformations are key for the experimental production of high-dimensional maximally entangled Bell-states.
\end{sciabstract}

\section*{Introduction}

The polarization of photons is a well-studied and reliable degree-of-freedom for the transmission of information. Using simple optical components such as half and quarter wave-plates, one can perform any unitary operation for polarization. However, photon polarization resides in a two-dimensional space, where the maximal information content of a single photon is limited to one bit. Having access to more than one bit per photon is not only conceptually interesting, but allows the implementation of novel advanced quantum communication and computation problems \cite{Araujo2014, Tavakoli2015}. For example, moving to a larger alphabet in quantum key distribution not only increases the key generation rate, but also provides improved resistance against noise and advanced eavesdropping attacks \cite{Bourennane:2002uo, HuberPawlowski:2013, Malik2014}.

There are several options for exploring discrete high-dimensional photonic degrees-of-freedom. For example, one can send a photon into one out of many possible paths \cite{Christoph, OBrien}. In such a ``path-encoding,'' it is also known how to perform arbitrary unitary transformations \cite{Reck}. However, path-encoding is not well suited for the purpose of communication due to very strict alignment and stability requirements. A more suitable degree-of-freedom is the spatial structure of photons, which involves Hermite-Gauss \cite{CartesianBeams}, Ince-Gauss \cite{Ince, InceEnt} or Laguerre-Gauss modes \cite{Allen1992, Mair2001} in the paraxial approximation. In particular, photons with a Laguerre-Gaussian mode structure can carry integer values $\ell$ of OAM with a helical phase front which goes from 0 to 2$\pi \ell$. As the OAM is theoretically unbounded, it gives access to a large state space. For this reason, it has been used in many classical \cite{Gibson, Willner, Mozart} and quantum communication experiments \cite{RochesterQKD, ItalyHallExperiment, ViennaEntanglement}, as well as in the investigation of quantum entanglement in large Hilbert spaces \cite{Vaziri, Dada, 100Dim}.

The ability to perform arbitrary unitary transformations directly in the OAM degree-of-freedom would greatly expand its use in quantum information. An indirect approach to carrying out such transformations is to transfer the information encoded in OAM to the path degree-of-freedom \cite{ModeSorter,RochesterSorter, OAMDM}, perform the transformation in the path-encoding and transfer the information back into OAM \cite{RobertInterface,HilbertHotel}. While this method is theoretically possible, it is technically difficult and suffers from the limitations imposed by the optical transformations involved. The question then arises whether there exist simpler and more direct ways for performing certain transformations of OAM modes. Unitary transformations that are useful for quantum information schemes often involve transformations between states in mutually unbiased bases (MUBs). For example, quantum cryptography with polarization requires transformations between states in the right/left-circular (R/L) and horizontal/vertical (H/V) bases. To go from the R/L to the H/V basis, one can perform a MUB transformation by using a quarter wave-plate and then perform a cyclic operation in the H/V basis (using a half-wave plate) to access both states. Similarly, transformations in the higher dimensional space of OAM can be broken down into transformations between high-dimensional mutually unbiased bases and cyclic operations within those bases \cite{Tavakoli2015}. Interestingly, an n-fold cyclic transformations is an $n^{\textrm{th}}$-root-of-unity transformation (as it fulfills the property $\hat{P}^n = \mathbb{I}$, where $\mathbb{I}$ is the identity), which performs a rotation in an n-dimensional space.  

Here we present an experimental implementation of a four-dimensional cyclic transformation of modes within the OAM basis. Our experimental scheme is able to perfectly (in a lossless manner) cycle through the 4-dimensional alphabet Furthermore, it only uses simple optical elements that manipulate the spatial mode of a light beam. An important building block of the experiment is an interferometer that was designed to sort photons based on their OAM content \cite{Leach2002}. Here we use it as a two-input, two-output OAM-parity beamsplitter, as implemented in a recent experiment \cite{Malik:2015we}. Interestingly, the experimental configuration for this cyclic transformation was found by a computer algorithm designed by some of the authors of this paper \cite{Krenn2015}. The algorithm uses a toolbox of experimentally available components, from which it creates experimental configurations. The property of the resulting states and transformations are then calculated and compared to a list of selection criteria. For the experiment presented here, the lossless performance of cyclic rotation in a high-dimensional state space was used as the criterion. If the experiment fulfills the criterion, it is automatically simplified and reported. Surprisingly, even though the final experiment consists of only a small number of components, the authors were not able to find an implementation by themselves. As a result, this setup is the first computer-designed quantum experiment that has been successfully implemented in the laboratory.  

\subsection*{Orbital Angular Momentum Beamsplitter}

\begin{figure}[h!]
\centering
	\centering
	\includegraphics[width=0.8\textwidth]{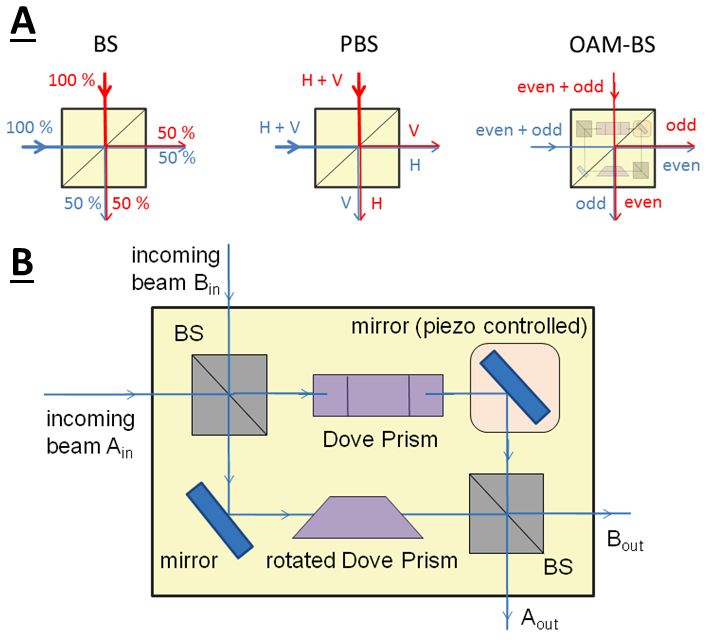}
\caption{\textbf{Orbital Angular Momentum beamsplitter (OAM-BS):} A) Comparison between a non-polarising beamsplitter (BS), a polarising beamsplitter (PBS) and an OAM-BS. B) The OAM-BS is a Mach-Zehnder interferometer with two Dove prisms, one in each path, oriented at an angle of $\pi/2$ relative to each other. This interferometer has the ability to sort even and odd values of OAM into its two output paths. One of the mirrors is piezo-controlled for fine adjustment.}
\label{fig:OAMBS}
\end{figure}

An essential building block of our setup is an orbital angular momentum beamsplitter (OAM-BS), which consists of a Mach-Zehnder interferometer (MZI) with an additional Dove prism in each arm (see Fig. \ref{fig:OAMBS}). In 2002, Leach et al. \cite{Leach2002} developed this interferometer to sort individual photons based on the parity of their OAM mode, in an analogous way to how a polarising beamsplitter (PBS) sorts individual photons based on their polarisation. Many such interferometers could then be cascaded to distinguish between arbitrarily many OAM modes with a theoretical efficiency of 100\%. When an OAM mode with an azimuthal phase structure $\textrm{e}^{i\ell\phi}$ is rotated by an angle $\alpha$, it acquires an additional phase given by $\textrm{e}^{i\ell\alpha}$. Notice that the phase acquired is proportional to both, the angle of rotation $\alpha$, as well as the OAM quantum number $\ell$ of the mode \cite{Courtial1998}. 

The Dove prisms in the MZI are rotated by an angle $\alpha/2$ with respect to each other. When $\alpha = \pi$ the relative phase difference between the two arms is $\ell\pi$, which only depends on the $\ell$ value of the incoming photons. As a result, photons entering the MZI from the incoming path A$_{in}$ with odd values of $\ell$ undergo constructive interference in arm A$_{out}$ and destructive interference in arm B$_{out}$. For photons from A$_{in}$ that carry an even value of $\ell$ the opposite happens. They go only into arm B$_{out}$ (with an additional sign flip) and never into path A$_{out}$ due to constructive and destructive interference, respectively. If the incoming path B$_{in}$ is used, the same happens (see Fig. 1 for a graphical representation).

\section*{Four-fold cyclic transformation for OAM}

\begin{figure}[h!]
\centering
\includegraphics[width=0.7\textwidth]{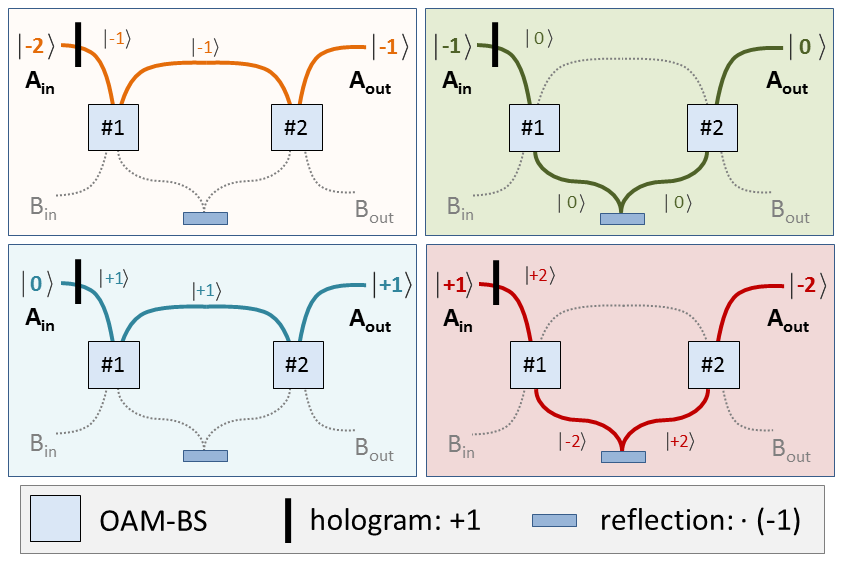}
\caption{\textbf{Conceptual sketch of the experimental setup.} The photons enter input $A_{in}$ and propagate to a spiral phase hologram (solid black rectangle) that adds one OAM quantum ($+1$) to the input mode. They then enter the first orbital angular momentum beamsplitter (OAM-BS) through the upper input arm (the lower input arm is not used).  The first OAM-BS sorts even and odd values of OAM into different paths. Modes coming out in the lower path undergo a sign change of the OAM value via one reflection on a mirror. The second OAM-BS sorts the modes again with only the upper output path returning the transformed modes in all cases (odd and even). The experiment is depicted four times in different colors for each possible input mode.}
\label{fig:setup_rough}
\end{figure}

The cyclic transformation introduced here is a unitary transformation that maps a particular OAM mode within an orthonormal set of modes to the next one in a fixed cycle. The four OAM modes that we use in our experiment are given by $\ell$ = $-2$, $-1$, $0$, and $+1$. To find an experimental implementation, we applied the computer algorithm MELVIN discussed in Ref.~\cite{Krenn2015} to search for high-dimensional cyclic rotations using only experimentally accessible linear optics components. One of the resulting configuration of a four-fold cycle requires one spiral phase hologram with $\ell=+1$, two orbital angular momentum beamsplitters (OAM-BS) and a reflection in one of the possible paths between them (see Fig.~\ref{fig:setup_rough}). In this setup, only the upper input arm ($\textrm{A}_{in}$) of the first OAM-BS is used and only the upper output arm of the second OAM-BS ($\textrm{A}_{out}$) will be needed. However, both paths between the two OAM-BSs are important for the cyclic transformation.

\begin{figure}[h]
\centering
\includegraphics[scale=0.6]{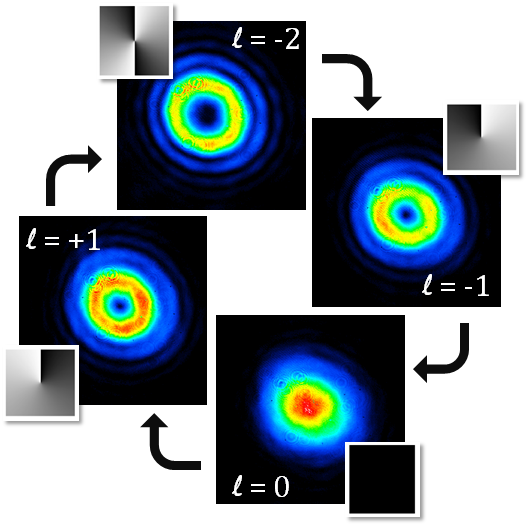}
\caption{\textbf{Visualization of the four-fold of OAM modes between $\ell$ = -2 and $\ell$ = 1.} False color images are recorded with a CCD camera in our experimental setup and the insets show their theoretically calculated phase front.}
\label{fig:cyclictrafo}
\end{figure}

We will now discuss in detail how the experimental implementation works. If the upper incoming beam ($\textrm{A}_{in}$) possesses OAM of $\ell$ = $-2$, it becomes $\ell$ = $-1$ after adding an OAM quantum of $+1$ with a spiral phase hologram. The first OAM-BS sorts this (odd) mode into the upper arm, from where it is input into the second OAM-BS. Once again, the OAM-BS sorts this odd mode into the upper output arm ($\textrm{A}_{out}$). The transformation from $\ell$ = $-2$ to $\ell$ = $-1$ is then complete (see fig.~\ref{fig:cyclictrafo})

In general, the experimental setup transforms the input $\ell_{in}$ into the output $\ell_{out}$ according to

\[ 
 \ell_{out} =
  \begin{cases}
   \ell_{in} + 1 	& . . . \: \ell \text{ even} \\
   -(\ell_{in} + 1)& . . . \: \ell \text{ odd}
  \end{cases}
\]

\begin{figure}[h]
\centering
\includegraphics[scale=0.6]{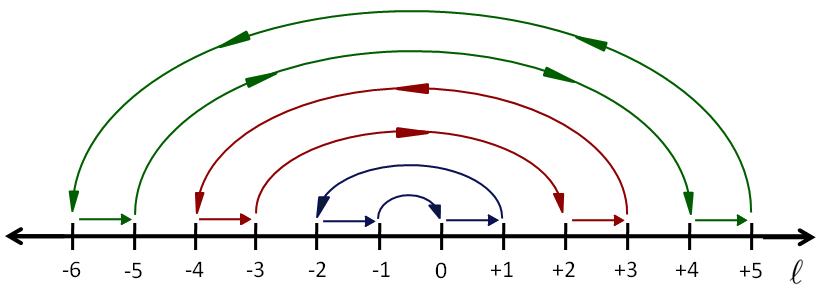}
\caption{\textbf{Four-fold transformations possible with the current setup.} Each closed loop of arrows shows the set of four OAM modes that can be cycled through with our experimental setup. In this paper the blue inner cycle is realized. However, in principle more four-fold cycles are accessible as can be seen in the outer arcs of red and green.}
\label{fig:trafo}
\end{figure}

\noindent According to this equation, higher-order four-fold cycles are realizable with the same setup. The general transformation behaviour for higher-order mode sets is visualized in Fig.~\ref{fig:trafo}, where the set of modes we experimentally investigate here ($\ell = -2, -1, 0, +1$) forms the lowest-order cycle.

It is important to distinguish between the transformation that involves adding OAM quanta to a particular mode, and the cyclic transformation discussed here. A single spiral phase plate or a spatial light modulator (SLM) can perform the transformation from $\ell_{in}$ = $-2$ to $\ell_{out}$ = $-1$, $\ell_{in}$ = $-1$ to a Gaussian beam with $\ell_{out}$ = $0$, or from the Gaussian input $\ell_{in}$ = $0$ to $\ell_{out}$ = $+1$. However, a single component does not comply with a cycle as can be seen in the final step: the input of $\ell_{in}$ = $+1$ is naturally mapped onto $\ell_{out}$ = $+2$ and not onto the desired $\ell_{out}$ = $-2$, as the loop property requires. These components only realize linear transformations and not cyclic ones. It is important to note that even though thin holograms (such as SLMs or spiral phase plates) can impart arbitrary phase shifts on the spatial mode, they cannot perform an arbitrary mode transformation if the impinging mode is unknown.

\section*{Experimental Details}

\begin{figure}[h!]
\centering
\includegraphics[scale=0.8]{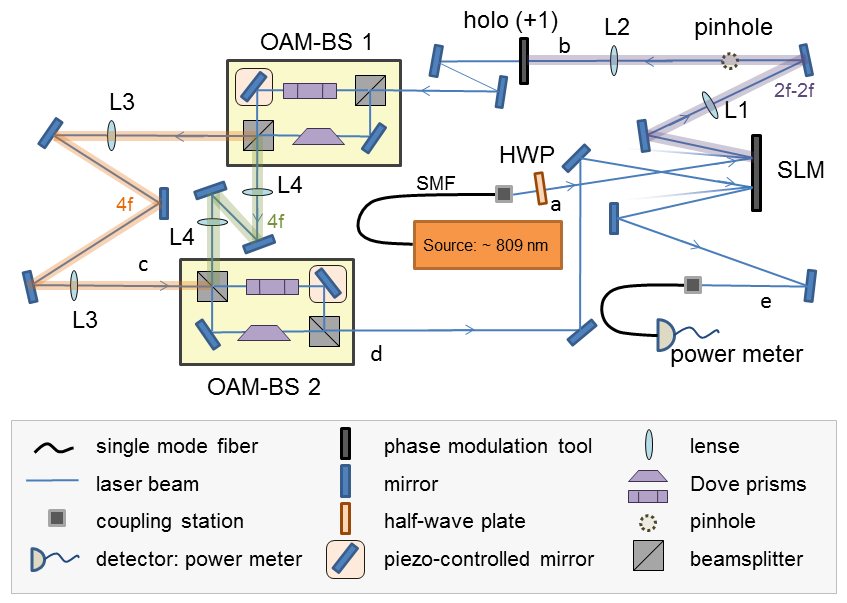}
\caption{\textbf{Schematic of the experimental setup.} The source is a grating-stabilized laser ($\lambda\approx809$nm). A single mode fiber (SMF) provides a  Gaussian mode that is horizontally polarized by a half-wave plate (HWP) and incident on a diffractive hologram displayed on a spatial light modulator (SLM). A 2$f$-2$f$ imaging system (L1 and L2) is used to demagnify the beam size and filter out the first diffraction order via a pinhole. After this preparation, the beam is modulated by an etched hologram of $\ell=+1$ and fed into the first orbital angular momentum beamsplitter (OAM-BS 1). Even input OAM modes are sorted by the OAM-BS into the longer path with a 4f-system of lenses (L3). Odd input modes are sent into the shorter path with another 4f-system of lenses (L4). Both of these arms are inputs of OAM-BS 2. After the OAM-BS 2, the cyclic transformation is complete. To measure the content of cycled modes we redirected the beam to the unused half of the SLM, where a hologram in combination with an SMF acts as an OAM mode filter. The power in the desired mode is measured with a power meter.}	
\label{fig:setup_whole}
\end{figure}

A schematic of the experimental setup can be seen in Fig.~\ref{fig:setup_whole}. The laser is an externally grating-stabilized laser diode centered at around 809nm. We use a single-mode fiber to spatially filter the laser beam, allowing us to send a collimated Gaussian mode ($\ell$=$0$) into the experimental setup. A half-wave plate is used to ensure the photons are horizontally polarised. This is necessary for proper modulation by a computer-controlled phase-only spatial light modulator (SLM, Holoeye-Pluto) that is used for generating the input OAM modes \cite{Jesacher2004, Gibson2004cv, OAMDM}. After the SLM, a 2$f$-2$f$ imaging system is used to filter out the modulated first diffraction order with a pinhole in the Fourier plane. Lenses with $f_{1} = 300 \: \text{mm}$ and $f_{2} = 150 \: \text{mm}$ demagnify the beam to half its size such that it fits through the etched hologram (holo +1) with a size of $4\times4\text{ mm}^2$.

One orbital angular momentum beamsplitter (OAM-BS) consists of two 50:50 beamsplitters, two Dove prisms rotated $\pi /2$ with respect to each other, and two mirrors. One mirror is mounted on a piezo-controlled stage in order to adjust the path length on a nanometer scale (see Fig.~\ref{fig:OAMBS}). Dependent on the parity of the OAM value of the incoming beam, one output arm of the first OAM-BS is occupied. An even input for instance, goes into the longer arm between the two OAM-BSs, where a 4$f$-system with plano-convex lenses ($f_{3} = 250 \text{mm}$) is implemented to avoid propagation effects from having a detrimental effect on the beam quality. Three mirrors flip the sign of the OAM mode. The shorter arm has a 4$f$-system of lenses L4 with $f_{4} = 100 \text{mm}$. The two 4$f$ systems also ensure that the inputs to OAM-BS 2 are conjugate (image planes) to the outputs of OAM-BS 1. The second OAM-BS is identical to the first one. However, both of its inputs are used, making its alignment more challenging. 

We measure the OAM content of the mode after each step of the the cyclic transformation via a projective measurement \cite{Mair2001}. A hologram displayed on the SLM can be used to ``flatten'' the phase of a specific incident OAM mode, which then couples efficiently to a single-mode fiber. Using a power meter, we can measure how much power was carried by the mode we are projecting into. Thus, the hologram and the single mode fiber together act as a mode filter. Since the display of our SLM is large enough to fit two different holograms, we use the left side for mode preparation and the right side for measurement. 

\section*{Experimental results and discussion}

\begin{figure}[h!]
\centering
\includegraphics[scale=0.7]{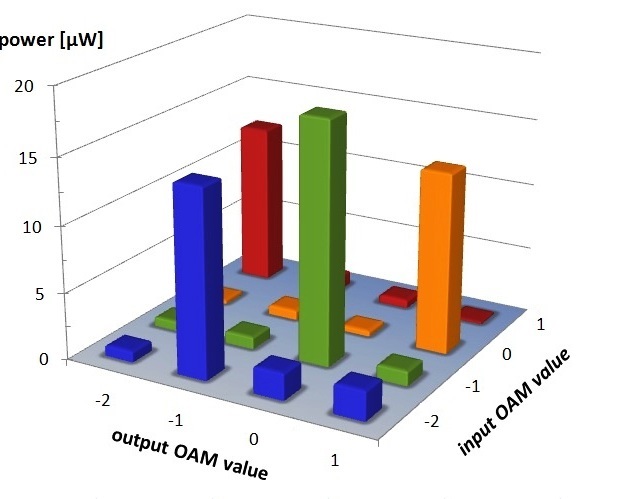}
\caption{Measured power in all four modes for every input $\ell$ value. Each colored row corresponds to measurements performed for one particular input OAM mode.}
\label{fig:CyclicOAM_graph}
\end{figure}

The experimental setup is tested by sending all four OAM modes in our set ($\ell=-2, -1, 0, +1$), one after the other. For each input mode, the mode content of the output is measured by performing projective measurements. If our cyclic transformation works, each input mode should be shifted to the adjacent mode in the cyclic set. The measured data showing the power in each output mode as a function of input mode number is plotted in Fig.~\ref{fig:CyclicOAM_graph}.

The four-fold cyclic transformation operation of our experimental setup is confirmed by the distinct peaks that appear in the correct output mode values, as shown in Fig.~\ref{fig:CyclicOAM_graph}. The efficiency $E$ of receiving the correct modes after the cyclic transformation of a given input mode are calculated in Table 1.

\begin{table}[h!]
\begin{center}
\begin{tabular}{c|cccc} 
\\
input mode $\ell$	& \textbf{-2}	& \textbf{-1}	& \textbf{0}	& \textbf{+1}	\\
\\
\hline
\\
$ E = I_{c} / I_{total} $ & 
(73.8 $\pm$ 1.0) \%   & 
(97.0 $\pm$ 1.0) \%   & 
(90.7 $\pm$ 1.5) \%   & 
(87.0 $\pm$ 1.5) \%   \\ \\
\end{tabular}
\end{center}
\caption{Measured efficiencies: For a given input mode, we show the probability that the correct mode is identified. $I_{c}$ is the intensity of the correctly identified mode, and $I_{total}=\sum_{l=-2}^{+1} I_l$ is the total intensity collected at the four output modes.}
\end{table}
From this data, it is clear that while our cyclic transformation works well, its operation is not uniform across the set of modes. Higher-order modes seem to be transformed with a lower probability than lower-order modes. This can be attributed to the larger diameter of higher-order modes that are clipped at certain optical elements---in particular at the $4\times4$mm$^2$ hologram. This problem can easily corrected by using smaller beams or larger optical elements. Additionally, higher-order modes are known to have a lower coupling efficiency in a phase-flattening measurement than lower-order modes \cite{Qassim}. The use of a mode-dependent lens on the SLM could be used to correct this behavior.

\section*{Conclusion and Outlook}

We have experimentally demonstrated a four-fold cyclic transformation of optical modes carrying orbital angular momentum (OAM). Modes from a set of OAM quantum numbers $\ell=-2,-1,0,+1$ are shifted into the adjacent mode in the set in a cyclic manner. Our setup offers theoretically lossless near-unit transformation efficiency that is reduced to an average efficiency of around 87\% due to technical limitations. The experiment was performed with commercially available optical components and without active interferometric stabilization. We expect the measured efficiency to approach the theoretical value with the specific improvements discussed above.

The four-dimensional cyclic transformation demonstrated here opens the door towards performing arbitrary mode transformations between high-dimensional mutually unbiased bases (MUBs)---a capability considered key for realizing quantum information systems in high dimensions. The experimental setup was found by using a computer algorithm (MELVIN) designed for finding high-dimensional transformations of quantum states \cite{Krenn2015}. The results serve as confirmation of MELVIN's ability to find experimental configurations that achieve a desired transformation.

While we have performed our experiment with a classical beam of light, it should be noted that the setup operation is identical for single photons. However, for its use in quantum experiments, it is important that this cyclic transformation also works for photons carrying a superposition of OAM modes. In order to do so, the two path lengths from the first OAM-BS to the second must be matched to within the coherence length of the input photon. The experiment would then consist of three cascaded Mach-Zehnder interferometers and would require additional improvements in stability. Interestingly, this experiment can be extended with a little effort to an eight-dimensional cyclic operation in a hybrid space of OAM and polarization \cite{Krenn2015}. The cyclic transformation demonstrated here shows great promise for applications in future quantum information systems and may prove invaluable in fundamental tests of quantum mechanics \cite{Lapkiewicz2011iq,DAmbrosio:2013di}.

\section*{Acknowledgement}
We thank Nora Tischler for help with the experiment. This project was supported by the European Research Council (SIQS Grant 600645 EU-FP7-ICT) and the Austrian Science Fund with SFB F40 (FOQUS). MM acknowledges funding from the European Commission through a Marie Curie fellowship (OAMGHZ).

\end{document}